\documentstyle[preprint,revtex,eqsecnum]{aps}
\tightenlines
\begin{document}
\begin{title}
Quasielastic neutrino scattering from oxygen \\
and the atmospheric neutrino problem
\end{title}
\author{J. Engel}
\begin{instit}
Bartol Research Institute, University of Delaware, Newark, DE  19713
\end{instit}
\author{E. Kolbe, K. Langanke}
\begin{instit}
W.K. Kellogg Radiation Laboratory, California Institute of
Technology, \\
Pasadena, CA  91125
\end{instit}
\author{P. Vogel}
\begin{instit}
Physics Department, California Institute of Technology,
Pasadena, CA  91125
\end{instit}
\begin{abstract}
We examine several phenomena beyond the scope of Fermi-gas models that
affect the quasielastic scattering (from oxygen) of neutrinos
in the 0.1 -- 3.0 GeV range.  These include Coulomb interactions of
outgoing protons and leptons, a realistic finite-volume mean field,
and the residual nucleon-nucleon interaction.
None of these effects are accurately
represented in the Monte Carlo simulations used to
predict event rates due to $\mu$
and $e$ neutrinos from cosmic-ray collisions in the atmosphere.  We
nevertheless conclude that the neglected physics cannot account for the
anomalous $\mu$ to $e$ ratio observed at Kamiokande and IMB, and is
unlikely to change absolute event rates by more than 10--15\%.  We
briefly mention other phenomena, still to be investigated in detail,
that may produce larger changes.
\end{abstract}
\pacs{}
\narrowtext

\section{Introduction}
\label{s:intro}

For some years now, an apparent anomaly has existed in the numbers
of $\mu$- and $e$-type neutrinos reaching the earth's surface after
being produced in the atmosphere by cosmic rays\cite{Kam,IMB}.
The observed ratio of muons to electrons
created in water \v{C}erenkov detectors is roughly 1:1 while
simple facts about the decay of pions and kaons in the
atmosphere lead one to expect a ratio much closer to 2:1.
Although it is difficult to predict absolute fluxes for each
kind of neutrino, errors tend to cancel in taking the ratio of the
two.  The roughly 2:1 expected ratio is robust, for example,
against the $\approx$ 20\% uncertainties in our knowledge of cosmic
ray fluxes and cross sections.  Furthermore,
any errors in calculating lepton production in the detector
are unlikely to affect the ratio because a cut can be
imposed on the momenta of the outgoing leptons.
If momenta are restricted to values significantly above the mass of
the muon, the cross sections for muon and electron production
ought to be nearly identical.

The scattering of atmospheric neutrinos from oxygen is nonetheless
worth investigating carefully.  Even if details in the structure
of ${}^{16}$O do not affect the $\mu$/$e$ ratio, they may alter
the total event rates considerably.  A substantial change would have
important consequences for the kinds of new physics that may be
responsible for the anomaly.  Monte Carlo simulations
using the calculated fluxes of Refs.\ \cite{Gaisser} and \cite{Honda}
imply that while roughly the
correct number of $e$ neutrinos are reaching the detector, far
fewer $\mu$ neutrinos are arriving than expected.
On the other hand, when the calculated fluxes of
Ref.\ \cite{Naum} are used the number of $\mu$
neutrinos appears to be correct, while too many
electrons and positrons are produced.
Attempts to resolve the problem\cite{th1,th2,th3} usually invoke the
conversion of $\mu$ neutrinos into $\tau$ neutrinos
on their way down from the top of the atmosphere.  If the treatment
of neutrino-oxygen scattering is not accurate, however,
this explanation might not be viable.
There may in fact be too many $e$ neutrinos, as well as a shortage
of $\mu$ neutrinos, no matter whose fluxes are correct; in that event
$\mu$-$e$ oscillations (or
some other new phenomenon like proton decay\cite{Mann}) would have to
play a role.  But theories incorporating these phenomena must
then avoid existing constraints e.g. from upward going
muons\cite{IMB2}
and solar neutrinos.
In this paper, we attempt to shed some light on the situation through
a careful examination of some aspects of the structure of
${}^{16}$O that affect charge-exchange cross sections for the
GeV-range atmospheric neutrinos.

The experiments are able to identify outgoing particles by the nature
of the \v{C}erenkov rings they produce.  By counting only
``single-ring'' events, experimenters can largely
restrict the data to charged-current
events in which electrons or muons are produced through quasielastic
scattering (collisions that produce pions usually result in more than
one ring).  At both Kamiokande and IMB, variants of the relativistic
Fermi-gas (RFG) model\cite{Moniz} are currently used to predict the
scattering cross sections (see, e.g., Ref.\ \cite{Naka}).  It is
not {\it a priori} obvious just how accurate the RFG model is in
this context; it works well, for instance, in
predicting electron scattering cross sections at certain energies
and angles\cite{Sick}, but fails to reproduce separated
longitudinal and transverse responses\cite{s1,s2,s3}.  A variety of
effects\cite{Van} can modify the ``free'' RFG response (which usually
mocks up binding effects through an average separation
energy\cite{Moniz}), and though
they are not all completely understood, they need to be examined as
thoroughly as possible in the context of neutrino scattering.  Here
we perform several calculations to test the role played by some of
the physical
effects not included in the Monte Carlos that simulate the behavior
of the detectors.
Usually we will assume nonrelativistic nuclear kinematics,
even though the incoming neutrinos can have energies up to a
few GeV (we offer some justification for this below).  To test
the importance of each new effect, we therefore compare our
results with those of the nonrelativistic Fermi-gas
(NRFG) model.  Fully relativistic nuclear models
exist\cite{QHD} and have been applied to quasielastic electron
scattering\cite{QHDQE1,QHDQE2}, but are more complicated and less
transparent than their nonrelativistic counterparts.  Our implicit
assumption is that
relativistic treatments of new effects will result in corrections of
the same order to the RFG as do ours to the NRFG.

The physics we assess includes the role played by bound states and
resonances, the Coulomb interactions of the outgoing
leptons and nucleons with the remaining A=15 nucleus, and finally the
residual two-body interaction between
nucleons in ${}^{16}$O.  The first
two of these require a finite-volume model, which we develop in the
next section.  Interactions can be included in this model and
(more schematically)
in nuclear matter; we carry out both calculations
in Sec.\ \ref{s:interaction}.  In Sec.\ \ref{s:discussion} we assess
the cumulative effect of our results on atmospheric neutrino rates.
Our conclusion is that while the new effects modify the rates to a degree
and introduce some uncertainty, they do not dramatically alter
the predictions of the RFG Monte Carlos.

\section{Fermi-gas models and finite-volume effects}
\label{s:volume}

The starting point for our investigation is the RFG model.
Here, nucleons are Dirac spinors occupying plane wave states
up to the nuclear Fermi momentum (for ${}^{16}$O
about 225 MeV/c), and carrying the same
weak currents as free nucleons.  Nuclear binding is simulated by
subtracting an average value $\overline{E}$ (about 27 MeV) from all
occupied states; an energy transfer $\omega$ of at least
$\overline{E}$ is required for any inelastic scattering.
The model was initially viewed as an unexpected success, in light of
its simplicity, because it gave beautiful fits to early
electron-scattering data\cite{Sick}.  As mentioned
above, its shortcomings emerged only when transverse and
longitudinal responses were measured separately\cite{s1,s2,s3}.
The lack of total success in reproducing separated responses
is significant in our context because the quasielastic neutrino
response has a substantially larger transverse to longitudinal
ratio than the $(e,e')$ response at similar energies.  The
dominance of the transverse response is due largely to
the axial current, the ``charge'' component  of which involves
only the small components of the nucleon spinors.

We begin our quantitative study by citing
Ref.\ \cite{Kuramoto}, where the RFG model was studied
and compared with other models for the scattering of
neutrinos with energy below 300 MeV.
One result of that work was a close agreement between
total cross sections in the RFG and
NRFG models, provided a semileptonic current-current
interaction expanded to order $(q/M)^3$ was used in the latter.
We have modified the treatment in Ref.\ \cite{Kuramoto} to include
effects of order $p/M$ in the current (these are quite small),
and in addition we use completely nonrelativistic kinematics
to facilitate later comparison with potential-model calculations.
We find, as shown in Table \ref{t:FG}, that even at the energies
important for atmospheric
neutrino scattering, the agreement between the RFG and NRFG (with
the same values for the Fermi momentum and average binding, and
folded with the Kamioka neutrino flux\cite{Gaisser})
remains reasonably good.  That this is not preposterous
can be seen from Fig.\ \ref{f:omega}, where the distribution
in energy transfer $\omega$ is displayed for representative
values of the outgoing lepton momentum.  Because of the steep
drop with energy of the atmospheric neutrino spectrum\cite{Gaisser}
a substantial part of
the scattering occurs at low enough $\omega$ so that a
nonrelativistic treatment makes qualitative sense.
As already noted, we will consequently examine
other effects in the context of nonrelativistic models.
We will continue however to use the usual relativistic dipole
nucleon form factors with the standard cut-off values
$M_V$ = 0.84 GeV, $M_A$ = 1.032 GeV, the usual CVC form
of the weak magnetism term, and the Goldberger-Triemann
relation for the induced pseudoscalar term.
In addition we use the neutrino fluxes of Ref.\ \cite{Gaisser}
everywhere below, and as a measure of event rates for a given
lepton momentum we employ and expression for the total ``yield"
given by
\begin{equation}
Y(p_{lep}) = \int_{E_{lep}}^{\infty}
\frac{{\rm d}\sigma (E_{\nu})}{{\rm d}p_{lep}} f(E_{\nu})~dE_{\nu}~,
\end{equation}
where $E_{lep} = E_{\nu} - \omega$ and $f(E_{\nu})$ is the
incoming neutrino (or antineutrino) flux.

The Fermi-gas model treats a nucleus like ${}^{16}$O as if it
were nuclear matter with a slightly reduced density.
Within such a context it is difficult to incorporate the
physics of bound states, resonances, and Coulomb repulsion,
which derive from the finite extent of the nucleus.  The steep drop
in the neutrino flux (reflected in Fig.\ \ref{f:omega}) ought
to enhance these features;
to begin to understand their role we
model  $^{16}$O as 8 protons and 8 neutrons occupying the lowest three
levels of a standard Woods-Saxon potential
\begin{equation}
V(r) =  -V_0 f(r) + V_{ls} \frac{1}{r}\frac{{\rm d}f}
{{\rm d}r} {\bf l \cdot s } ~,
\label{e:WS}
\end{equation}
where
\begin{equation}
f(r) = \{ 1 + {\rm exp}[r - R]/a \}^{-1} ~,
\end{equation}
and $R, a$ are measures of the nuclear radius and diffuseness.
The final states consist of both bound and continuum eigenstates
of the same potential.  Corrections to the weak current up to
to order $(q/M)^3,~p/M$ are included in the same way as in our
version of the NRFG model.  A similar picture of neutrino scattering
was presented several years ago\cite{Bugaev}, but did
not consider the effects of lepton mass, or use a current with
corrections beyond 1/M.

Unfortunately the finite-well model described above will not necessarily
yield more accurate results than the Fermi-gas model.  As is shown
in Ref.\ \cite{Rosenfelder}, the average excitation energy at
a given momentum transfer $q$ in
a local potential (which must also be spin-independent --- not
the case here) must be the same as in the Fermi-gas model with
{\it no} binding added.  A shift can only arise from a non-local
potential or a two-body interaction.  This is not entirely surprising
because a local, energy-independent potential is not an accurate
representation of the mean nuclear field.  A better model is the
optical potential, which is known to be energy dependent or,
equivalently, non-local.  To some extent the average binding energy
inserted in the Fermi-gas model simulates the effects of two-body
interactions or of a non-local mean field.

We nonetheless use the simple single-particle potential model outlined
above as a starting point to which all kinds of additional physics
can be added.
To indicate the relation between this picture and the NRFG, we show
the ratio of yields in the two models in the first half of
Table\ \ref{t:WS}. Our Wood-Saxon
potential is specified by:  $V_0$=51 MeV, $V_{ls}$=32.8
MeV$\times$fm$^2$, $R$=1.27$\times A^{1/3}$, $a$=0.65 fm.
As shown in Table\ \ref{t:WS}, for $p_{lep} \leq$ 550 MeV/c
the potential model leads to slightly higher yields than the
NRFG.   We attribute this to the Wood-Saxon single-particle bound
states and resonances, the effects of which should be most
pronounced at low lepton momentum.

To incorporate the proton-nucleus Coulomb
interaction, we alter the potential
felt by outgoing protons by adding a repulsive interaction
associated with a spherical volume of
uniformly distributed charge.  As pointed out in
Ref.\ \cite{Donnelly},
the use of different interactions for initial and final states spoils
CVC; the magnitude of the problem, however, is small.
The second part of Table\ \ref{t:WS} shows the effect of Coulomb
repulsion of the outgoing protons, which are produced only by
incoming neutrinos (i.e. not by antineutrinos).
In accord with intuition, the repulsive
Coulomb interaction reduces their yield. When the contribution of
neutrinos and antineutrinos are added, however,
the magnitude of
the reduction, shown in the table, is only about 3\%.

Within the Wood-Saxon model we can also examine
the Coulomb interaction of the outgoing
charged lepton with the nucleus. It is tempting to treat the repulsion
as is commonly done for nuclear beta decay, i.e. by
multiplying the cross section by a Fermi function $F(Z,p)$
\begin{equation}
F(Z,p) = \frac {|\psi_{Coul}(r=R)|^2}{|\psi_{plane~wave}(r=R)|^2}
{}~ \rightarrow ~ \frac{\pm 2 \pi Z \alpha}
{1 - \exp(\mp 2 \pi Z \alpha)} ~,
\end{equation}
where $Z$ is the nuclear charge, $\alpha$ is the fine structure
constant, $R$ is the nuclear radius, and
the limit is for ultrarelativistic electrons and muons
($p/E \rightarrow 1$).  This procedure, however, is
valid only for outgoing $s$-wave particles; it clearly cannot be
applicable more generally since the ultrarelativistic limit is not
unity, as it ought to be.
In our case, because $pR$ can be much greater than 1,
we need a better procedure.  Guided
by the distorted-wave picture applied in the analysis of
quasielastic electron scattering\cite{Rosenfelder}, we
replace the momentum of the charged outgoing
lepton by an effective value
\begin{equation}
p_{eff} = p ( 1 + \frac{\langle V \rangle}{E}) ~,
\langle V \rangle = \pm \frac {4 Z \alpha }{3R} ~
\end{equation}
when evaluating the nuclear matrix element. ($\langle V \rangle$ is
the mean value of the Coulomb potential inside the nucleus.)
The Coulomb correction treated in this way turns out to be very small,
on the order of 1\%.

\section{The residual interaction}
\label{s:interaction}

So far the models we have considered do not explicitly incorporate
residual two-body interactions between nucleons.  The Fermi-gas
models, with the extra 27 MeV binding, simulate their effects to
some degree, but it is not obvious whether a more realistic
treatment of the interactions will change event rates significantly.
In this section, we examine this question first for illustration
in the context
of nuclear matter --- that is as explicit two-body corrections
to the no-binding NRFG --- and then more rigorously in the
finite-volume model.
In both instances we will use the one tried and (reasonably)
true method for calculating continuum response: the Random
Phase Approximation (RPA).

In nuclear matter the calculation is
straightforward\cite{Fetter}.  The quasielastic response is
related to the particle-hole ``polarization'' propagator in medium,
which can be approximately evaluated as a sum of ring diagrams.
For the relevant part of the interaction, we use the standard
$\pi + \rho + \delta$-function
parameterization of the two-body particle-hole potential,
i.e. (in momentum space)
\begin{eqnarray}
\label{e:force}
V_{ph} & = & \left(f_0 '  +  V_{\pi} + V_{\rho} \right)
{\bf \tau}_1 \cdot {\bf \tau}_2 ~, \\
V_{\pi} & = & J_{\pi}(\omega,q) \left[ g_0 ' + {q^2 \over {\omega^2
- q^2- m_{\pi}^2} } \right] {\bf \sigma}_1 \cdot \hat{\bf q}
{\bf \sigma}_2 \cdot \hat{\bf q} ~, \nonumber \\
V_{\rho} & = & J_{\pi}(\omega,q) \left[ g_0 ' + { J_{\rho}(\omega,q)
\over {J_{\pi}(\omega,q)} } {q^2 \over {\omega^2
- q^2- m_{\rho}^2} } \right] ({\bf \sigma}_1 \times \hat{\bf q}) \cdot
({\bf \sigma}_2 \times \hat{\bf q} ) ~, \nonumber
\end{eqnarray}
where $f_0' = .6 / m_{\pi}^2$, $g_0' = .7$, and
\begin{equation}
\label{e:J}
J_{\pi} = 4 \pi {f^2_{\pi N N} \over m_{\pi}^2 }
F_{\pi}^2(\omega,q) ~,~
J_{\rho} =  4 \pi {f^2_{\rho N N} \over m_{\rho}^2 }
F_{\rho}^2(\omega,q) ~.
\end{equation}
The $\pi$ and $\rho$ coupling strengths $f_{\pi N N}$ and
$f_{\rho N N}$,
and the form factors $F_{\pi}$ and $F_{\rho}$ are defined in
Ref.\ \cite{Osterfeld}, which also contains many relevant references.
The spin-singlet force is a pure (Landau-Migdal)
contact force, while in the spin triplet channel, the
two terms correspond to $\pi$ and $\rho$ exchange supplemented by a
phenomenological contact force (softened by pion form factor) to
approximately account
for short range correlations.  Though we have used
couplings that appear commonly in the literature, no consensus
exists on the values these parameters take in medium; a simultaneous
reproduction of high $q$ electron-scattering and
$p-p$ data has so far proved elusive.  We will therefore also consider
as an alternative a pure density-dependent Landau-Migdal interaction
commonly used to describe
low-energy excitations in finite nuclei\cite{Co}.

Fig.\ \ref{f:rings} shows cross sections for production of
electrons as a function of energy loss $\omega$ for a fixed
neutrino energy of 600 MeV.  Here one can see explicitly how
the 27 MeV binding added to the NRFG simulates two-body interactions;
both cause the strength to be depleted (compared to the NRFG with
no binding) at low energies and enhanced at high energies.  The
size of this effect is comparable when 27 MeV binding is added
to the NRFG and when the force of Eq.\ (\ref{e:force}) is
used.  On the other hand, the pure contact force, which is probably
less realistic, depletes
considerably more strength from the low-$\omega$ region, and adds
only a tiny amount at very high $\omega$.  The same pattern is
present for other neutrino energies as well.

The shift in strength
is not hard to understand.  The cross section is dominated by
the transverse response, in which the pion plays no role.  The
$\rho$-exchange piece of Eq.\ (\ref{e:force}), which {\it is} active,
is repulsive up to values of $q$ around 770 MeV and hardens
the transverse response. The Landau-Migdal force remains repulsive
for all values of $q$ and therefore has an even
larger effect. These features are
reflected in the event rates calculated with the various models,
shown in Table \ref{t:matter}. Since
the atmospheric neutrino spectrum weights relatively low values of
$\omega$, the event rates are reduced by a shift of strength to
higher excitation energies.
The reduction caused by the ``standard'' force is nearly exactly equal
to that induced by the 27 MeV shift of the free Fermi-gas response,
while the pure Landau-Migdal
interaction yields rates that are about 12\% smaller.  Although
the matter of what kind of force to use at such high values
of $q$ and $\omega$ is still not resolved\cite{Brown}, the
Landau-Migdal interaction is near the edge of plausibility, and in
nuclear matter we can
probably take the spread of values to represent the maximum
uncertainty due to our ignorance of nuclear forces in medium.

Even with a perfect force, however, a nuclear matter calculation has
to be viewed as schematic.  We therefore discuss effects of the
residual interaction in a finite volume as well.  Here we use
an implementation of the continuum RPA described in Ref.\
\cite{Buba} and applied to lower-energy neutrino scattering in
Ref.\ \cite{Kolbe}.  In this approach the basic building blocks are
coherent superpositions of continuum creation and
bound-state annihilation operators and their hermitian
conjugates.  Integro-differential RPA equations for these phonons
can be derived and solved, yielding explicit expressions for
ground state correlation and transition amplitudes.
Technical problems arising from finite range forces are solved by
an expansion in Weinberg states\cite{Raw}.

Here also, we use two distinct forces to gauge the uncertainty in our
calculation.  The first is the same pure Landau-Migdal interaction
defined in Ref.\ \cite{Co} and used above (though because we are now
working in a finite volume, density-dependent terms that have
no effect in nuclear matter come into play).
The second interaction, of finite range, is a parameterization of
the G-matrix associated with the Bonn
meson-exchange potential\cite{Machl}.  Fig.\ \ref{f:response} shows
cross sections for the two forces, alongside the free Wood-Saxon
response, for neutrinos of 600 MeV, as in Fig.\ \ref{f:rings}.
A similar though not identical result emerges.  The
Landau-Migdal interaction
again depletes the low-energy region and enhances the response
very slightly at high energy.  The Bonn potential, however, has
little effect on the cross section.  [This result is consistent
with other studies\cite{Osterfeld}, which show that G-matrix
calculations tend to yield somewhat less hard-core repulsion than
Landau-Migdal parameterizations of low-energy data]
Event rates with the two forces are shown in Table \ref{t:residual}.
(We were not able to satisfactorily apply our RPA code
above $p_{lep}$ of 500 MeV/c.)
The similarities between the entries of Tables
\ref{t:matter} and \ref{t:residual} suggest that the magnitude
of the effect does not depend sensitively on the
size of the nuclear volume.
\section{Discussion}
\label{s:discussion}

Before summarizing our findings, we must note that
several effects not included in this work may alter cross sections
and bear investigating.  The configurations mixed into the wave
functions by our RPA treatment are of the nucleon-hole type.  But
virtual delta-hole excitations can also admix; the role they
play in our process is not yet known.  In addition, not all
single-ring events need come from quasielastic scattering.  Some
occur, for instance, when pions are created below \v{C}erenkov
threshold, or when two nucleons are ejected from the nucleus.
The first effect is included in some approximation in the
experimental
Monte Carlo codes, but two-nucleon knockout is completely ignored.
There is reason\cite{exchange,Rost} to think that one or both
of these processes is
responsible for excess strength  between the quasielastic
peak and the delta-knockout region observed in electron
scattering\cite{s1}.  The underlying theory
has been developed in several ways, and while
no consensus exists on the details, the various methods
should be applied to this problem; at the very least, an
additional uncertainty in the rates can be estimated.

Another open problem is the possible modification of nucleon
form factors in medium\cite{Goe1,Brown}.
Again, there is no consensus as to the theoretical foundation
for such effects, and little empirical evidence
for them in weak processes. However, if the masses $M_A$ and $M_V$
in the vector and axial-vector form factors
are reduced by 10-15\% as suggested in the cited papers,
the cross section for both channels would be reduced by $\approx$
20\% for all values of $p_{lep}$ we consider.

Apart from all of this, our results in what we consider the most
realistic  model --- the Wood-Saxon well with all corrections and the
G-matrix based force --- differ only by a few percent
from the predictions
of the Fermi-gas model for the most important charged-lepton momenta
$p_{lep} \leq$ 550 MeV/c.  In part the size of the difference is due
to a cancellation between, for instance, bound-state and Coulomb
effects.  In any case, our most important finding is that none
of the effects we have considered can possibly alter
the $\mu/e$ ratio; in Tables I-IV the muon and
electron columns are nearly always identical. Our results support
the contention that the ratio is a robust measure of the anomaly.

The surprising agreement of our calculated absolute rates with
those of the Fermi-gas model carries some uncertainty; the
Landau-Migdal interaction reduces the yields
by about 13\%.  An application of that force to the transverse
(e,e') response\cite{Buba}, however,
leads us to to suspect that it
underestimates quasielastic neutrino cross sections.
In conclusion, then, our results generally support the Fermi-gas model
cross sections for purely
quasielastic processes in the momentum range relevant to the
atmospheric neutrino problem. Whether effects such as
two-nucleon knockout or excess pion production
increase the cross sections noticeably remains to be seen.
\acknowledgments

We wish to acknowledge useful discussions with E. Beier, G. Brown,
F. Cavanna, M. Ericson, T.K. Gaisser, S. Krewald, and T. Stanev.
This work was supported in part by National Science Foundation
under grants Nos. PHY91-08011, PHY90-13248, and PHY91-40397 and by
Contract No. DE-F603-88ER-40397 with the U.S. Department of Energy.

\newpage
\begin{table}
\caption{ Ratios of the lepton yields calculated in the
RFG and NRFG models, and the absolute ($\mu^-+\mu^+$) and
($e^-+e^+$) yields in units of $10^{-41}$(s$\cdot$MeV)$^{-1}$
in the RFG model.
}
\label{t:FG}
\begin{tabular}{c||llll|cc|cc}
$p$ (MeV/c) & $\mu^-$ & $\mu^+$ & $e^-$ & $e^+$
& $\mu^-+\mu^+$ & $e^-+e^+$ & yield ($\mu$) & yield($e$) \\
\tableline
150. & 0.95 & 1.05 & 0.94 & 1.02 & 0.96 & 0.95 & 6.80 & 4.29 \\
250. & 1.06 & 1.03 & 1.05 & 1.01 & 1.06 & 1.04 & 9.35 & 5.56 \\
350. & 1.15 & 1.01 & 1.14 & 1.00 & 1.13 & 1.12 & 7.40 & 4.12 \\
450. & 1.15 & 1.00 & 1.15 & 0.99 & 1.12 & 1.11 & 5.37 & 2.85 \\
550. & 1.14 & 0.98 & 1.13 & 0.97 & 1.10 & 1.10 & 3.97 & 2.07 \\
650. & 1.12 & 0.96 & 1.12 & 0.96 & 1.08 & 1.07 & 3.02 & 1.55 \\
750. & 1.10 & 0.94 & 1.10 & 0.93 & 1.05 & 1.05 & 2.35 & 1.20 \\
850. & 1.07 & 0.91 & 1.08 & 0.91 & 1.03 & 1.03 & 1.87 & 0.94 \\
950. & 1.06 & 0.90 & 1.07 & 0.90 & 1.01 & 1.02 & 1.51 & 0.76 \\
\end{tabular}
\end{table}

\begin{table}
\caption{
The yield ratios of $\mu^-+\mu^+$ and $e^-+e^+$
between the one-body Wood-Saxon-potential model and the NRFG model
(columns 2 and 3) and, within the Wood-Saxon model, the ratio of
yields with and without Coulomb interactions (columns 4 and 5).
}
\label{t:WS}
\begin{tabular}{c||cc|cc}
$p$ (MeV/c) & \multicolumn{2}{c|}{NRFG/WS}
& \multicolumn{2}{c}{with/without Coulomb} \\
& & & \multicolumn{2}{c}{for protons in WS} \\
\tableline
& $\mu^-+\mu^+$ & $e^-+e^+$ &  $\mu^-+\mu^+$ & $e^-+e^+$  \\
\tableline
150. & 1.05 & 1.04 & 0.95 & 0.95  \\
250. & 1.10 & 1.10 & 0.97 & 0.97  \\
350. & 1.09 & 1.09 & 0.97 & 0.97  \\
450. & 1.04 & 1.04 & 0.97 & 0.97  \\
550. & 1.02 & 1.01 & 0.97 & 0.97  \\
650. & 0.99 & 0.99 & 0.97 & 0.97  \\
750. & 0.96 & 0.96 & 0.97 & 0.97  \\
850. & 0.94 & 0.95 & 0.97 & 0.97  \\
950. & 0.93 & 0.94 & 0.98 & 0.98  \\
\end{tabular}
\end{table}

\begin{table}
\caption{
The yield ratios of $\mu^-+\mu^+$ and $e^-+e^+$
between the nuclear matter calculation (see text) and the NRFG
with binding.
Columns 2 and 3 were calculated with the Landau-Migdal force,
and columns 4 and 5 with the $\pi$ and $\rho$
exchange + $\delta$-function force \cite{Osterfeld}.
}
\label{t:matter}
\begin{tabular}{c||cc|cc}
$p$ (MeV/c) & \multicolumn{2}{c|}{nucl. matter/NRFG}
& \multicolumn{2}{c}{nucl. matter/NRFG} \\
& \multicolumn{2}{c|}{Landau-Migdal}
& \multicolumn{2}{c}{$\pi$+$\rho$ exch + $\delta$ } \\
\tableline
& $\mu^-+\mu^+$ & $e^-+e^+$ &  $\mu^-+\mu^+$ & $e^-+e^+$  \\
\tableline
150. & 0.87 & 0.85 & 0.86 & 0.84  \\
250. & 0.89 & 0.89 & 0.99 & 0.98  \\
350. & 0.90 & 0.89 & 1.02 & 1.02  \\
450. & 0.88 & 0.89 & 1.00 & 1.00  \\
550. & 0.88 & 0.88 & 0.98 & 0.98  \\
650. & 0.88 & 0.88 & 0.98 & 0.97  \\
750. & 0.88 & 0.87 & 0.96 & 0.96  \\
850. & 0.87 & 0.87 & 0.96 & 0.96  \\
950. & 0.87 & 0.87 & 0.95 & 0.95  \\
\end{tabular}
\end{table}

\begin{table}
\caption{
The yield ratios of $\mu^-+\mu^+$ and $e^-+e^+$
between the continuum RPA and the free response ( i.e., independent
nucleons in the Wood-Saxon potential).
Columns 2 and 3 were calculated with the Landau-Migdal force,
and columns 4 and 5 with the Bonn meson-exchange potential.
({\it Note that the bin centers are different here.})
}
\label{t:residual}
\begin{tabular}{c||cc|cc}
$p$ (MeV/c) & \multicolumn{2}{c|}{cont.RPA/free resp.}
& \multicolumn{2}{c}{cont.RPA/free resp.} \\
& \multicolumn{2}{c|}{Landau-Migdal} & \multicolumn{2}{c}{Bonn. pot.} \\
\tableline
& $\mu^-+\mu^+$ & $e^-+e^+$ &  $\mu^-+\mu^+$ & $e^-+e^+$  \\
\tableline
100. & 0.99 & 1.07 & 0.98 & 1.04  \\
200. & 0.86 & 0.87 & 0.93 & 0.92  \\
300. & 0.84 & 0.84 & 0.97 & 0.96  \\
400. & 0.85 & 0.85 & 0.98 & 0.98  \\
500. & 0.86 & 0.87 & 0.99 & 0.99  \\
\end{tabular}
\end{table}

\newpage
\figure{
Contribution in the RFG model of different energy transfers
$\omega$ to the
$\mu^-+\mu^+$ yield for muon momenta of 250 MeV/c (solid curve)
and 550 MeV/c (dashed curve).
\label{f:omega}}

\figure{
The response of the $^{16}$O nucleus to $\nu_e$'s of 600 MeV in
a reduced-density nuclear-matter calculation.
The differential cross section
is plotted vs. the energy transfer $\omega$.
The dot-dashed curve represents the NRFG response calculated
without the binding energy correction ($\overline{E} = 0$),
the dashed curve is the standard NRFG response
($\overline{E} = 27$ MeV), the dotted curve is the
response with the Landau-Migdal contact potential, and the
solid curve is response for the $\pi + \rho$
exchange + $\delta$-function force.
\label{f:rings}}

\figure{
The response of the $^{16}$O nucleus to $\nu_e$'s of 600 MeV in
a finite-volume calculation.
The differential cross section
is plotted vs. the energy transfer $\omega$.
The dashed curve represents the free response (independent nucleons),
the dotted curve
is the response with the Landau-Migdal contact potential, and
the solid curve represents the continuum RPA response calculated
with the Bonn meson-exchange G-matrix.
\label{f:response}}

\end{document}